\documentclass[journal=jpccck,manuscript=article]{achemso} 

\usepackage{textgreek}
\usepackage{amsmath}
\usepackage{graphicx}
\usepackage{siunitx}
\usepackage[version=3]{mhchem}

\author{Michal Horák}
\affiliation[CEITEC]
{Brno University of Technology, Central European Institute of Technology, Purky\v{n}ova 123, 612 00, Brno, Czech Republic}
\email{michal.horak2@ceitec.vutbr.cz}

\author{Michael Foltýn}
\affiliation[CEITEC]
{Brno University of Technology, Central European Institute of Technology, Purky\v{n}ova 123, 612 00, Brno, Czech Republic}

\author{Viktor Bajo}
\affiliation[CEITEC]
{Brno University of Technology, Central European Institute of Technology, Purky\v{n}ova 123, 612 00, Brno, Czech Republic}
\alsoaffiliation[UFI]
{Brno University of Technology, Faculty of Mechanical Engineering, Institute of Physical Engineering, Technická 2, 616 69, Brno, Czech Republic}

\author{Petr Dub}
\affiliation[CEITEC]
{Brno University of Technology, Central European Institute of Technology, Purky\v{n}ova 123, 612 00, Brno, Czech Republic}
\alsoaffiliation[UFI]
{Brno University of Technology, Faculty of Mechanical Engineering, Institute of Physical Engineering, Technická 2, 616 69, Brno, Czech Republic}

\author{Tomáš Šikola}
\affiliation[CEITEC]
{Brno University of Technology, Central European Institute of Technology, Purky\v{n}ova 123, 612 00, Brno, Czech Republic}
\alsoaffiliation[UFI]
{Brno University of Technology, Faculty of Mechanical Engineering, Institute of Physical Engineering, Technická 2, 616 69, Brno, Czech Republic}

\title{Plasmonics of non-noble metals}

\begin{document}

\begin{abstract}

Localized surface plasmon resonances are self-sustained, collective oscillations of free electrons in metallic nanostructures. They have a wide range of applications. The most common plasmonic metals are noble metals, such as gold and silver. However, there are applications, such as surface-enhanced Raman spectroscopy, in which using non-noble metals is advantageous. This review summarizes the investigation of localized surface plasmons in non-noble metal nanoparticles, providing an overview of the plasmonic properties of non-noble metals. We cover the following metals: aluminium (Al), antimony (Sb), bismuth (Bi), chromium (Cr), copper (Cu), gallium (Ga), indium (In), lead (Pb), magnesium (Mg), molybdenum (Mo), nickel (Ni), potassium (K), selenium (Se), sodium (Na), tellurium (Te), tin (Sn), titanium (Ti), tungsten (W), and zinc (Zn). Our summary therefore compares the plasmonic properties of non-noble metals and briefly introduces their potential to the readers.

\end{abstract}

\section{Introduction}

The interaction of the electromagnetic field and free electrons in metals at the metal-dielectric interface gives rise to hybrid light-matter states called surface plasmon polaritons (SPPs).  Their generation and propagation are studied by a branch of nanooptics called plasmonics. It is often sufficient to view the SPPs as an evanescent electromagnetic wave propagating along the metal-dielectric interface and exponentially decaying in the direction perpendicular to the interface.  In metallic nanostructures, collective oscillations of free electrons strongly couple to the electromagnetic field forming the excitations called localized surface plasmon resonances (LSPRs)\cite{10.1007/0-387-37825-1}. The clear mathematical foundation for LSPRs was established by Gustav Mie in 1908 \cite{10.1002/andp.19083300302}. Their characteristic feature is a strong enhancement of the electromagnetic field within the surrounding dielectric together with its confinement on the subwavelength scale, which can be utilized to control various optical processes in a wide spectral region even below the free space diffraction limit \cite{10.1038/nphoton.2010.237, 10.1038/nmat2630}. The significance of this feature is further increased by the easy tunability of the optical properties of nanostructures by engineering their size, shape, or dielectric environment \cite{10.1021/jp026731y}. This flexibility in design allows one to create a large amount of optical functions. Therefore, plasmonic nanostructures have a wide field of applications \cite{10.1088/2040-8986/aaa114}, including photonics \cite{10.1126/science.1203056}, microscopy \cite{10.1103/physrevlett.94.057401}, spectroscopy \cite{10.1146/annurev.physchem.58.032806.104607}, energy harvesting \cite{10.1016/j.mattod.2013.09.003}, medicine \cite{10.1002/lpor.201200003}, sensing \cite{10.1039/c2cc33266c, 10.1021/nl500574n}, and catalysis \cite{10.1021/acs.nanolett.7b04776, 10.1021/acs.accounts.9b00224, 10.1038/s41467-025-57569-0}.

New discoveries with high application potential are often connected to the implementation of new concepts in the field of plasmonics, such as a plasmoelectric effect \cite{10.1126/science.1258405}, plasmonic lasing \cite{10.1038/nphoton.2008.82}, generalized laws of reflection \cite{10.1126/science.1210713}, spin-orbit coupling \cite{10.1038/nphoton.2015.232}, chirality \cite{10.1002/adma.201205178}, or Babinet's principle of complementarity \cite{10.1103/physrevlett.93.197401, 10.1103/physrevb.76.033407, 10.1038/s41598-019-40500-1}. In addition, unconventional plasmonic materials are utilized in specific application fields, including the prospect of spectro-electrochemistry of silver amalgam nanoparticles \cite{10.1021/acs.jpcc.9b04124} or tunable plasmonic devices or metasurfaces made of phase-changing materials such as vanadium dioxide \cite{10.1021/acsphotonics.5b00249, 10.1021/acsphotonics.1c00222, 10.1515/nanoph-2023-0824, 10.1021/acsnano.4c13188} or gallium \cite{10.1002/adom.201900307, 10.1515/nanoph-2020-0314, 10.1088/1361-6463/abaae2}. Additionally, the native oxide layer on non-noble metallic nanostructures can suppress the catalytic activity of the nanoparticles similarly to the activity achieved by covering gold nanoislands with a graphene cover layer \cite{10.1073/pnas.1205478109} or by shell-isolated nanoparticles \cite{10.1021/acs.jpclett.6b00147}. 

The most common plasmonic metals are gold, silver, and to a lesser extent aluminium. They were found to be ideal plasmonic materials for their availability, relative chemical stability, nontoxicity, and ease of production in the industry. The only issue regarding aluminium and silver is the instability of antennas due to oxidation, which with time changes the structure and, with it, the plasmonic behavior. Furthermore, these metals (except aluminium) can be used for plasmonic applications only in the visible and near-infrared parts of the spectrum \cite{10.1021/jp405773p}, as their performance is restricted at lower wavelengths by interband transitions. Consequently, gold supports LSPR at wavelengths longer than \SI{550}{\nano\meter} and silver supports LSPR above \SI{350}{\nano\meter}. Therefore, it is vital to explore other plasmonic materials, which can be used in visible to near-ultraviolet spectral region. Metals are generally considered good plasmonic materials in a certain spectral region if their real part of the dielectric function is negative and their imaginary part of the dielectric function, related to the losses, is small enough. The ultraviolet and whole visible spectral range is covered by aluminium \cite{10.1021/nn405495q}, bismuth \cite{10.1021/jp3065882, 10.1021/acs.jpclett.5c02531}, gallium \cite{10.1021/nn5072254, 10.1021/acs.jpclett.3c00094, 10.1021/acs.jpclett.5c02035}, magnesium \cite{10.1021/acs.nanolett.8b00955} and silver amalgam \cite{10.1021/acs.jpcc.9b04124}. The ultraviolet plasmonic activity for chromium, copper, indium, lead, palladium, platinum, rhodium, ruthenium, tellurium, tin, titanium and tungsten was also theoretically studied and discussed \cite{10.1021/jp405773p, 10.1039/c3cp43856b, 10.1039/c5cp90112j}. The aim of this review is to summarize the knowledge in the field of plasmonics of non-noble metals and inspire future investigations in this field.

\section{Noble and non-noble metals}

A noble metal is a metallic chemical element that is generally resistant to corrosion and oxidation, even at high temperatures. It is usually found in nature in its raw form. The most common noble metals include gold (Au), silver (Ag), and platinum group metals: platinum (Pt), palladium (Pd), rhodium (Rh), ruthenium (Ru), osmium (Os), and iridium (Ir). In more specialized fields of applications, the number of elements counted as noble metals can be smaller or larger. However, we will follow this division. Plasmonic activity for all noble metals has been explored, including osmium \cite{10.1016/j.jscs.2023.101651} and iridium \cite{10.1021/jacs.8b05105, 10.1016/j.apsusc.2019.04.268} that have not been mentioned in the Introduction. 

\begin{figure}[t]
\centering
  \includegraphics[width=8cm]{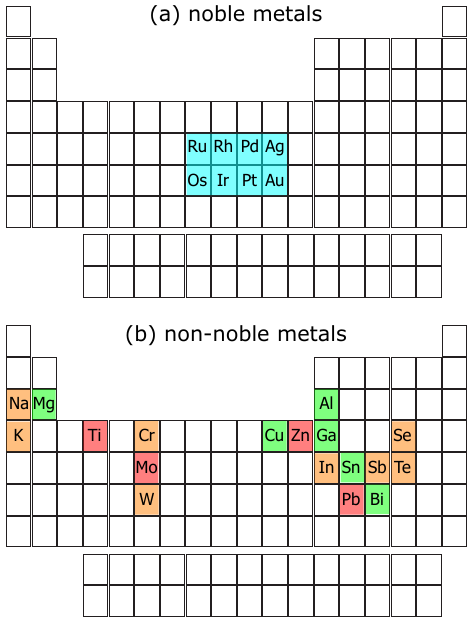}
  \caption{Noble (a) and non-noble (b) metals used in plasmonic applications. In the following, we focus mostly of the non-noble metals marked by green color as they are commonly used in plasmonic applications, whereas the non-noble metals marked by orange color are less employed in plasmonic applications. In addition, plasmonic activity of non-noble metals marked by red color was just predicted with no reported experimetal applications.}
  \label{Fig1}
\end{figure}

Non-noble metals are metals that are generally not resistant to corrosion and oxidation, especially in moist or acidic environments. They tend to react more easily with other substances, such as oxygen or acids, compared to noble metals, which are more stable and inert. In this review, we focus on non-noble metals, i.e., all metals that are not listed above in the list of noble metals. The following non-noble metals were studied and discussed: aluminium (Al), antimony (Sb), bismuth (Bi), chromium (Cr), copper (Cu), gallium (Ga), indium (In), lead (Pb), magnesium (Mg), molybdenum (Mo), nickel (Ni), potassium (K), selenium (Se), sodium (Na), tellurium (Te), tin (Sn), titanium (Ti), tungsten (W), and zinc (Zn). In the following, we summarize the research in the field of plasmonic properties of these non-noble metals in detail. Figure \ref{Fig1} shows a periodic table with marked noble and non-noble metals used in plasmonic applications.

The following discussion focuses on an overview of the investigations of the plasmonic properties of non-noble metal nanostructures, as well as the key parameters for plasmonics, such as their experimental dielectric functions (some of them were obtained using Refractiveindex.info \cite{10.1038/s41597-023-02898-2}), the theoretical quality factor of LSPR considering the figure of merit derived from the dielectric function as $Q_\mathrm{LSPR}=-\Re{(\epsilon)}/\Im{(\epsilon)}$, and the Fröhlich energy, i.e. the energy at which the Fröhlich condition is fulfilled \cite{10.1021/jp405773p}. It is important to note that the Fröhlich condition is the condition for LSPR in metallic spherical nanoparticles much smaller than the wavelength of light embedded in a homogeneous medium. This condition is derived from the electrostatic approximation of the Mie theory when the denominator of the polarizability expression is close to zero. This condition gives rise to a pronounced oscillation of free electrons, a substantial enhancement of the electric field in proximity to the nanoparticle, and elevated levels of light absorption and scattering at a particular resonant frequency. This resonant frequency corresponds to a value at which the real part of the nanoparticle's permittivity approximates $-2$ times the real part of the surrounding medium's permittivity. In the context of a metallic nanoparticle with dilectric function $\epsilon$ in air, the Fröhlich condition is $\Re[\epsilon(E_\mathrm{F})]=-2$. The value of $E_\mathrm{F}$ then indicates the theoretical energy of the LSPRs in a small nanosphere in the air. The energy of LSPRs can be further tuned by the dielectric environment that embeds the nanostructure and by enlargement of the nanostructure and its morphology. Table \ref{Tab1} summarizes the Fröhlich energy $E_\mathrm{F}$ for selected non-noble metals.

\begin{table}[t]
\small
\centering
  \caption{Fröhlich energy of selected non-noble metals}
  \label{Tab1}
  \begin{tabular}{|c|c|c|}
    \hline
    material & $E_\mathrm{F}$ [\SI{}{\electronvolt}]  & dielectric function reference\\
    \hline
    Al & 8.9 & Rakić \cite{10.1364/AO.34.004755} \\
    Bi & 5.7 & Werner et al. \cite{10.1063/1.3243762} \\
    Cu & 3.4 & Babar \& Weaver \cite{10.1364/AO.54.000477} \\
    Ga & 8.5 & McMahon et al. \cite{10.1039/c3cp43856b} \\
    Mg & 6.4 & Hagemann et al. \cite{10.1364/JOSA.65.000742} \\
    Sn & 10.6 & Palik \cite{10.1016/c2009-0-20920-2} \\
    \hline
  \end{tabular}
\end{table}

\section{Aluminium}

\begin{figure*}[ht]
 \centering
 \includegraphics[width=16cm]{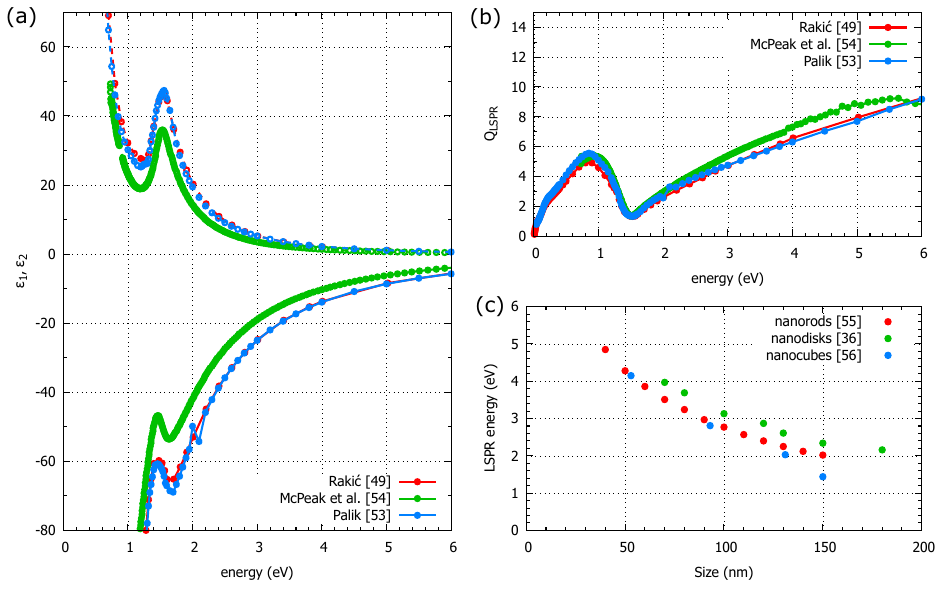}
 \caption{Aluminium plasmonics: (a) Experimental dielectric functions of aluminium by Rakić  \cite{10.1364/AO.34.004755}, McPeak et al. \cite{10.1021/ph5004237}, and Palik \cite{10.1016/c2009-0-20920-2}. The real part $(\epsilon_1)$ is plotted by filled circles connected with a solid line and the imaginary part $(\epsilon_2)$ by empty circles connected with a dashed line. (b) Theoretical quality factors of LSPRs derived from these dielectric functions as $Q_\mathrm{LSPR}=-\epsilon_1/\epsilon_2$. (c) LSPR energy as a function of size of aluminium nanostructures, namely longitudinal dipole mode in aluminium nanorods \cite{10.1021/nl303517v}, in-plane dipole mode in aluminium nanodisks \cite{10.1021/nn405495q}, and dipole mode in aluminium nanocubes \cite{10.1021/acsnano.9b05277}.}
 \label{Fig_aluminium}
\end{figure*}

Aluminium is the 12th most abundant element in the universe. It is a soft, nonmagnetic, ductile, and low density metal with a melting temperature of \SI{660.3}{\celsius}. It is a CMOS compatible material. It has a great affinity for oxygen and a protective oxide layer (\ce{Al2O3}) on the surface that forms rapidly when exposed to air, stabilizing at 2.5--\SI{3}{\nano\meter} thickness in hours and acts as a self-limiting passivation layer preventing further oxidation for at least 30 days \cite{10.1021/nl080453i}. Its biocompatibility is generally good. However, for direct contact with the body, surface treatments, such as anodizing, are crucial to enhance corrosion resistance and prevent adverse reactions by creating a protective oxide layer. Aluminium has emerged as a promising plasmonic material characterized by a high bulk plasma frequency of approximately 15 eV, resulting from the high electron density with 3 conduction electrons per atom. The negative real part of the dielectric function extends well into the ultraviolet range, allowing LSPR to reach the deep ultraviolet through visible to near-infrared regions of the spectrum. However, it has a spectrally localized narrow-band interband transition around \SI{1.5}{\electronvolt} that may lead to a strong interaction with LSPR resulting in hybridization \cite{10.1021/nl400328x}.

Figure \ref{Fig_aluminium}a shows a comparison of three experimental dielectric functions of aluminium available in the literature, namely by Rakić  \cite{10.1364/AO.34.004755}, McPeak et al. \cite{10.1021/ph5004237}, and Palik \cite{10.1016/c2009-0-20920-2}. There are no significant differences. The peak in the imaginary part of the dielectric function around \SI{1.5}{\electronvolt} corresponds to the narrow-band interband transition. The real part of the dielectric function reaches highly negative values, promising a good plasmonic activity over the entire energy range from the ultraviolet to the infrared spectral region. Figure \ref{Fig_aluminium}b shows the theoretical quality factors of the LSPRs derived from these dielectric functions as $Q_\mathrm{LSPR}=-\epsilon_1/\epsilon_2$. There are no significant differences between the theoretical quality factors derived from the three examples of dielectric functions. It reaches values between 2 and 6 in the near-infrared, 2 to 6 in the visible, and 6 to 10 in the ultraviolet spectral region. The lowest value of 2 in the red part of the spectrum is attributed to the narrow-band interband transition around \SI{1.5}{\electronvolt}.

Aluminium nanostructures can be manufactured using various litography techniques, such as electron beam lithography, nanosphere lithography, nanoimprint lithography, and photolithography, or by chemical synthesis of nanoparticles in various geometries \cite{10.1088/0022-3727/48/18/184002, 10.1021/acsnano.9b05277, 10.1021/acs.nanolett.7b04820}. LSPR in various types of aluminium nanostructures have been studied, such as nanorods \cite{10.1021/nl303517v, 10.1021/acs.nanolett.7b04820}, nanodisks \cite{10.1021/nn405495q}, nanodisk arrays \cite{10.1021/nl501460x}, and nanocubes \cite{10.1021/acsnano.9b05277}, demonstrating their wide tunability from ultraviolet through the visible to near-infrared spectral region. The influence of the oxide layer on the LSPR in aluminium nanostructures was studied, resulting in a redshift in the resonant frequency and a decrease in the scattering efficiency with an increasing oxide fraction \cite{10.1021/nn405495q, 10.1039/c7nr04904h}. aluminium nanostructures can degrade by photocorrosion under strong ultraviolet light exposure that may be prevented by depositing a protective layer \SI{5}{\nano\meter} \ce{TiO2} or a protective layer \SI{10}{\nano\meter} \ce{SiO2} \cite{10.1021/acs.jpclett.9b02137}.

Figure \ref{Fig_aluminium}c shows the dipole LSPR energy as a function of the size of aluminium nanostructures for three representative systems available in the literature, namely the longitudinal dipole mode in nanorods \cite{10.1021/nl303517v}, the in-plane dipole mode in nanodisks \cite{10.1021/nn405495q}, and the dipole mode in nanocubes \cite{10.1021/acsnano.9b05277}. The results show a wide tunability of aluminium nanostructures from the ultraviolet region for the size of structures below \SI{100}{\nano\meter} to the visible spectral region for the size of structures up to \SI{200}{\nano\meter}. The structures above \SI{200}{\nano\meter} are intended to represent the plasmonic platform for the near-infrared spectral region.

There are numerous applications for the plasmonic properties of aluminium nanostructures, including surface-enhanced Raman spectroscopy \cite{10.1063/1.4746747}, plasmon-enhanced fluorescence \cite{10.1021/jp2122714, 10.1021/ac900263k}, surface-enhanced infrared absorption \cite{10.1515/nanoph-2023-0131}, and biosensing \cite{10.1021/am404509f}. In addition, a wide field of applications for such nanoparticles in photocatalysis was introduced, including water oxidation \cite{10.1021/acsnano.6b04004}, methylene blue decomposition \cite{10.1007/s11356-022-22861-9}, hydrogen dissociation \cite{10.1021/acs.nanolett.5b05149}, and the reverse Boudouard reaction, which is an endothermic heterogeneous process occurring at high temperatures where carbon dioxide reacts with solid carbon to produce high-purity carbon monoxide \cite{10.1038/s41563-020-00851-x}. Consequently, aluminium is a well established plasmonic material with deep-explored plasmonic properties that is used in a wide range of applications.

\section{Bismuth}

\begin{figure*}[t]
 \centering
 \includegraphics[width=16cm]{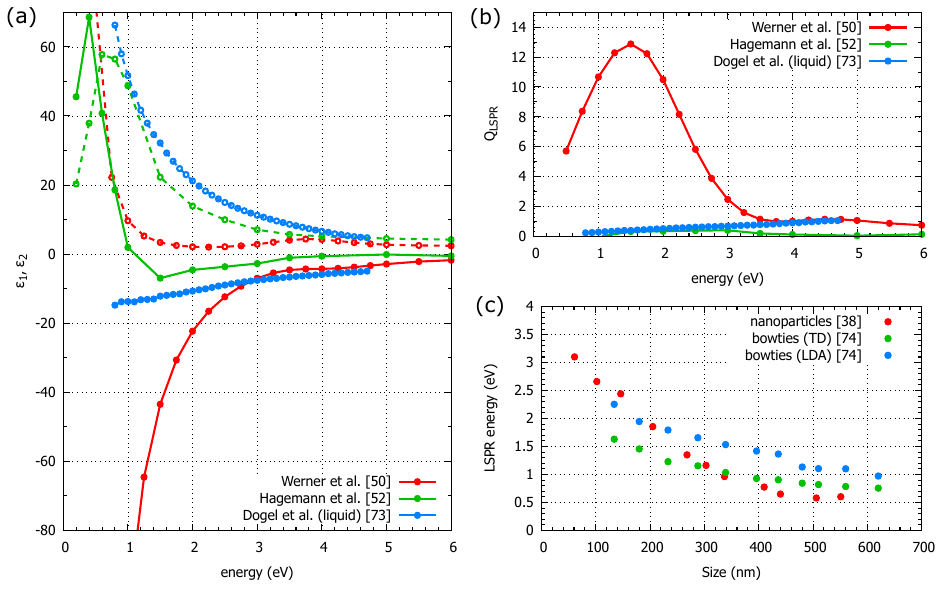}
 \caption{Bismuth plasmonics: (a) Experimental dielectric functions of bismuth by Werner et al.  \cite{10.1063/1.3243762}, Hagemann et al. \cite{10.1364/JOSA.65.000742}, and for the liquid bismuth by Dogel et al. \cite{10.1103/PhysRevB.72.085403}. The real part $(\epsilon_1)$ is plotted by filled circles connected with a solid line and the imaginary part $(\epsilon_2)$ by empty circles connected with a dashed line. (b) Theoretical quality factors of LSPRs derived from these dielectric functions as $Q_\mathrm{LSPR}=-\epsilon_1/\epsilon_2$. (c) LSPR energy as a function of size of bismuth nanostructures, namely dipole mode in bismuth nanoparticles on silicon dioxide membrane \cite{10.1021/acs.jpclett.5c02531}, transverse dipole (TD) and longitudinal antibonding dipole (LDA) mode in bowties on sillicon nitride membrane manufactured by focused ion beam lithography \cite{10.1021/acsnano.5c07482}.}
 \label{Fig_bismuth}
\end{figure*}

Bulk bismuth is a metal with a melting temperature of \SI{271.5}{\celsius}. It is generally considered biocompatible. The only known bodily harm caused by bismuth is inflammation of the lungs after inhalation of fine bismuth powder, which is caused by mechanical irritation of the tissue. However, when bismuth is introduced into the body in other forms, no negative effects on mammals have been observed, even at doses as high as \SI{1000}{\milli\gram} per \SI{1}{\kilo\gram} of body mass \cite{10.1539/joh.47.242, 10.1039/D0CS00031K}. Bismuth in the form of thin polycrystalline layers appears to be chemically stable and resistant to oxidation under ambient conditions; however, bismuth can easily be oxidized in an oxygen-reactive atmosphere, such as oxygen plasma, or by annealing under an oxygen atmosphere \cite{10.1021/acsnano.5c07482}. As a result, bismuth can be easily covered by a few nanometers of oxide layer that may act as an insulator and, for example, prevent charge transfer in catalytic reactions. The low effective mass of free electrons and the dielectric function of bismuth suggest that it is an attractive plasmonic material suitable for plasmonics spanning from the near-infrared to the ultraviolet spectral region \cite{10.1103/PhysRevLett.98.076603, 10.1155/2018/3250932, 10.1039/c3cp43856b}. Furthermore, the extraordinary properties of bismuth, including quantum confinement \cite{10.1063/1.2192624}, temperature-induced metal-to-semiconductor transition \cite{10.1088/0957-4484/21/40/405701, 10.1016/j.tsf.2025.140678}, and high values of the Seebeck coefficient \cite{10.1002/anie.201005023}, when combined with its plasmonic performance, have the potential to yield new applications. The biocompatibility, high atomic number, and accessible functionalization of bismuth nanoparticles make them a compelling candidate as a contrast enhancing agent in medical X-ray imaging and computational tomography techniques \cite{10.1016/j.jddst.2021.102895, 10.1016/j.jpcs.2018.03.034}.  Furthermore, their recently reported photocatalytic and photothermal properties also make them attractive for use in photothermal cancer therapies \cite{10.1021/acsanm.7b00255, 10.1039/D0CS00031K}, environmental remediation \cite{10.1039/D3EN00983A, 10.1021/jacs.3c04727, 10.1039/C4CC02724H}, and energy storage \cite{10.1016/j.ensm.2023.03.023}.

Figure \ref{Fig_bismuth}a shows a comparison of three experimental dielectric functions of bismuth available in the literature, namely by Werner et al.  \cite{10.1063/1.3243762}, Hagemann et al. \cite{10.1364/JOSA.65.000742}, and for the liquid bismuth by Dogel et al. \cite{10.1103/PhysRevB.72.085403}. There are significant differences in these dielectric functions. The dielectric function of Werner et al. \cite{10.1063/1.3243762} indicates that bismuth is a suitable plasmonic material, whereas the other two indicate that it is not. Figure \ref{Fig_bismuth}b shows the theoretical quality factors of the LSPRs derived from these dielectric functions. Taking the dielectric function of Werner et al. \cite{10.1063/1.3243762}, bismuth represents a good plasmonic platform for the near-infrared and visible spectral region with the theoretical quality factor of the LSPRs reaching values from 2 to 13. In contrast, the use of the other two dielectric functions results in the theoretical quality factor of the LSPRs below 1. Therefore, it is necessary to select a suitable dielectric function for this material when considering its plasmonic applications.

The potential combination of plasmonic applications and the extraordinary properties of bismuth has driven research on circular or spherical bismuth nanostructures \cite{10.1021/acs.inorgchem.1c02621, 10.1016/j.photonics.2022.101058, 10.1021/jp3065882, 10.1364/ol.45.000686, 10.1002/adom.202302130, 10.1021/acs.jpclett.5c02531}, bismuth nanowires \cite{10.1063/1.1337940, 10.1103/PhysRevB.65.115418}, and nanostructured bismuth thin films \cite{10.1016/j.photonics.2022.101058, 10.1021/jp3065882, 10.1002/adom.202302130, 10.1039/C7CP04359G, 10.1021/acsnano.5c07482}. Most studies on the plasmonic response of chemically synthesized bismuth nanoparticles have been limited only to investigations of plasmonic performance using far-field optical spectroscopy. The main constraint of this method is that it mostly measures the overall response of an ensemble of nanoparticles in a large volume of solution containing nanoparticles of various sizes resulting in a difficult isolation of a single nanoparticle contribution. The exploration of the spectral tunability of LSPRs in individual bismuth nanostructures as a function of their size has recently been investigated by electron energy loss spectroscopy for monocrystalline spherical bismuth nanoparticles \cite{10.1021/acs.jpclett.5c02531}, as well as bismuth plasmonic antennas manufactured by focused ion beam lithography \cite{10.1021/acsnano.5c07482}, and the theoretically predicted spectral interval that extends from the near-infrared to the ultraviolet was confirmed for LSPRs in bismuth nanostructures.

Figure \ref{Fig_bismuth}c shows the dipole LSPR energy as a function of the size of bismuth nanostructures for two representative systems available in the literature, namely the dipole mode in bismuth nanoparticles on the silicon dioxide membrane \cite{10.1021/acs.jpclett.5c02531}, and the transverse dipole (TD) and longitudinal antibonding dipole (LDA) mode in bowties on the sillicon nitride membrane manufactured by focused ion beam lithography \cite{10.1021/acsnano.5c07482}. The results show a wide tunability of bismuth nanostructures from the ultraviolet region for the size of structures below \SI{50}{\nano\meter} through the visible spectral region for the size of structures up to 200--\SI{300}{\nano\meter} depending on the shape to the near-infrared spectral region for larger structures.

In addition, bismuth has been considered as a viable substitute for gold. A direct comparison of the LSPR energies in bismuth and gold bowties suggests that bismuth can be considered as an alternative material to gold \cite{10.1021/acsnano.5c07482}. The Q factors of the bismuth antennas are marginally lower than those of gold; however, this disadvantage is counterbalanced by their consistent performance even at higher plasmon energies. Further research in bismuth-based plasmonics could include thin-film optimization to enhance the properties of bismuth antennas. In addition, a deeper understanding of the metal-to-semiconductor transition in very thin bismuth layers has the potential to be applicable in the low-temperature active plasmonic devices. In the case of nanoparticles, the main open question is their possible functionalization, which is crucial for any application in biochemistry.

\section{Copper}

\begin{figure*}[ht]
 \centering
 \includegraphics[width=16cm]{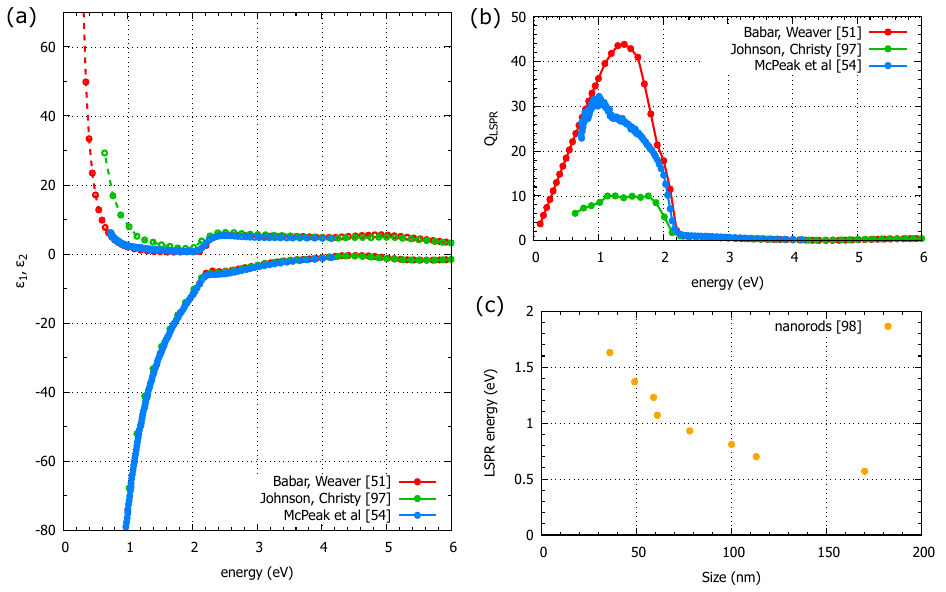}
 \caption{Copper plasmonics: (a) Experimental dielectric functions of copper by Babar and Weaver \cite{10.1364/AO.54.000477}, Johnson and Christy \cite{10.1103/PhysRevB.6.4370}, and McPeak et al. \cite{10.1021/ph5004237}. The real part $(\epsilon_1)$ is plotted by filled circles connected with a solid line and the imaginary part $(\epsilon_2)$ by empty circles connected with a dashed line. (b) Theoretical quality factors of LSPRs derived from these dielectric functions as $Q_\mathrm{LSPR}=-\epsilon_1/\epsilon_2$. (c) LSPR energy as a function of size of copper nanoparticles, namely longitudinal dipole mode in copper nanorods dispersed in tetrachloroethylene \cite{10.1021/acs.nanolett.0c02648}.}
 \label{Fig_copper}
\end{figure*}

Copper is a soft, malleable, and ductile metal with very high thermal and electrical conductivity. Its melting temperature is \SI{1084.6}{\celsius}. It is a CMOS compatible material. Copper is an indispensable trace metal element in the human body, as it being an important component and catalytic agent of many enzymes and proteins, while copper-based materials generally exhibit antibacterial effects and good biocompatibility \cite{10.1016/j.bioactmat.2020.09.017}. Copper does not react inherently with water; however, it slowly reacts with atmospheric oxygen, resulting in the formation of a layer of brown-black copper oxide. This layer functions as a protective barrier that protects the underlying metal from further corrosion. A green layer of copper carbonate is often visible on aged copper structures. Copper tarnishes when exposed to sulfur compounds, with which it reacts to form various copper sulfides. Fortunately, copper nanostructures can be protected from harsh environments by protective coatings, such as thin oxide films, including alumina (\ce{Al2O3}), hafnia (\ce{HfO2}), or titania (\ce{TiO2}), deposited by atomic layer deposition onto substrate-based copper nanostructures. The resulting nanostructures are resistant to oxidation, high temperatures, and aqueous, acidic, and alkaline solutions without unduly influencing important plasmonic properties \cite{10.1021/acs.chemmater.0c02715}. 

Figure \ref{Fig_copper}a shows a comparison of three experimental dielectric functions of bismuth available in the literature, namely by Babar and Weaver \cite{10.1364/AO.54.000477}, Johnson and Christy \cite{10.1103/PhysRevB.6.4370}, and McPeak et al. \cite{10.1021/ph5004237}. There seem to be no significant differences. The real part of the dielectric function reaches highly negative values in the infrared and visible spectral range, promising good plasmonic activity, especially in the infrared spectral region. The imaginary part of the dielectric function increases around \SI{2.1}{\electronvolt} due to the interband transition. Figure \ref{Fig_copper}b shows the theoretical quality factors of the LSPRs derived from these dielectric functions. Taking the dielectric function of Babar and Weaver \cite{10.1364/AO.54.000477}, copper represents a perfect plasmonic platform for the near-infrared and red visible spectral region (below \SI{2}{\electronvolt}) with an excellent theoretical quality factor of LSPRs that reach values of 20 to 45. The use of the other two dielectric functions results in lower quality factors, but still reaching very high values of 10 to 30 in the near-infrared spectral region. Therefore, copper is supposed to be an excellent plasmonic platform for the near-infrared and visible spectral region.

Copper nanoparticles of various shapes can be synthesized using multiple strategies, including polyol synthesis with effective control over their morphology \cite{10.1016/j.apsusc.2024.161640}, disproportionation reaction in oleylamine solvent, providing shape-selective synthesis of spherical and cubic nanocrystals with a very narrow size distribution \cite{10.1021/jp5014187}, as well as colloidal synthesis \cite{10.1021/accountsmr.2c00134} or plasmon-driven synthesis \cite{10.1021/acs.nanolett.5c02359}. Although localized surface plasmons in copper are strongly damped by the interband transition above \SI{2.1}{\electronvolt} \cite{10.1021/acs.chemmater.5b03519}, they are comparable to gold in a low loss window extending from 1.6 to \SI{2.0}{\electronvolt} \cite{10.1021/nl070648a}. Finally, elongated copper nanostructures sustain spectrally sharp visible to infrared plasmons of a quality factor similar to that of their silver counterparts; therefore, copper is emerging as an attractive, cheap and abundant material platform for high-quality plasmonics in elongated nanostructures \cite{10.1021/acs.nanolett.0c04667}. The longitudinal dipole mode of localized surface plasmons in copper nanorods was proved to be tunable from 0.6 to \SI{1.6}{\electronvolt} by their aspect ratio \cite{10.1021/acs.nanolett.0c02648}. 

Figure \ref{Fig_copper}c shows the longitudinal dipole mode of LSPRs in copper nanorods \cite{10.1021/acs.nanolett.0c02648} that cover mostly the near-infrared spectral region with nanostructure size between 30 and \SI{300}{\nano\meter}. The larger structures are intended to represent the plasmonic platform for the infrared spectral region. Consequently, copper is supposed to be a promising plasmonic material in the red part of the visible and infrared spectral regions.

Nanostructured copper has been demonstrated to allow the conversion of carbon dioxide (\ce{CO2}) into a variety of reducing products by electrocatalysis \cite{10.1002/smtd.201800121}. Copper nanoparticles with intense visible region surface plasmon absorption bands were found to be excellent nonlinear scatterers \cite{10.1016/S0009-2614(02)00407-4}. Their applications include refractive index sensing, surface-enhanced Raman spectroscopy, single nanoparticle spectroscopy, or selective catalysis \cite{10.1039/c4nr04719b}. Surface-enhanced Raman spectroscopy of rhodamine B and photocatalysis in the paranitrophenol reduction reaction was demonstrated for hexagonal and cubic copper nanostructures \cite{10.1016/j.apsusc.2024.161640}. Copper nanoparticles were found to be an affordable plasmonic heater for photothermal applications including solar-vapor generation and driving temperature-dependent color changes in thermochromic molecules \cite{10.1039/d4sc04806g}. They were used in an interfacial solar steam generation system that performs solar desalination of water \cite{10.1038/s41598-023-40060-5}. Furthermore, they improve the performance of organic photovoltaics \cite{10.1063/5.0010427}. In addition, the plasmonic properties of copper oxides (\ce{Cu2O} and \ce{Cu3O2}) \cite{10.1088/1367-2630/ac219f} and copper monosulfide (CuS) \cite{10.15587/2706-5448.2021.237269} have been discussed. Further research in copper-based plasmonics could include thin-film optimization to explore and optimize the properties of copper antennas fabricated by lithography.

\section{Gallium}

\begin{figure*}[ht]
 \centering
 \includegraphics[width=16cm]{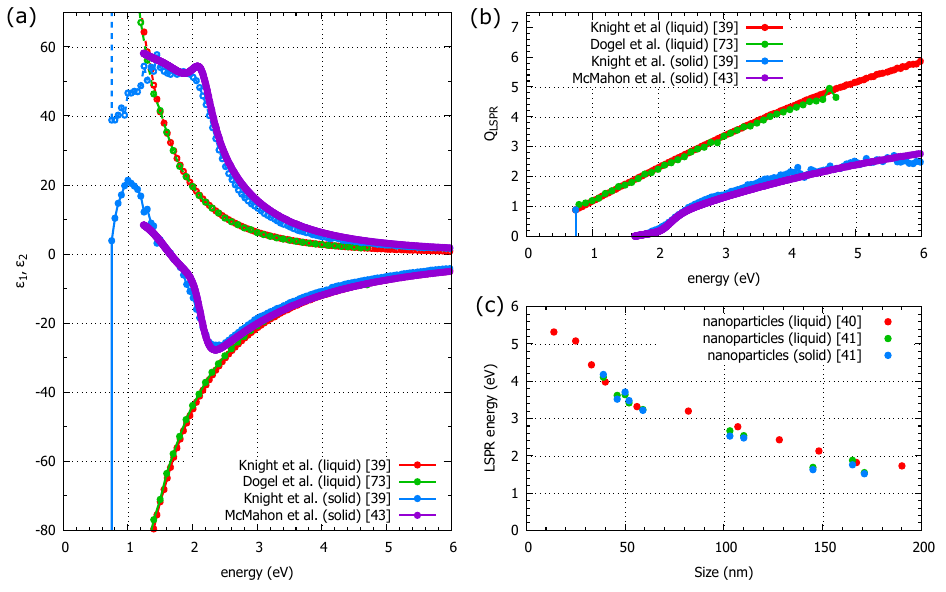}
 \caption{Gallium plasmonics: (a) Experimental dielectric functions of liquid gallium by Knight et al. \cite{10.1021/nn5072254} and Dogel et al. \cite{10.1103/PhysRevB.72.085403}, and solid gallium by Knight et al. \cite{10.1021/nn5072254} and McMahon et al. \cite{10.1039/c3cp43856b}. The real part $(\epsilon_1)$ is plotted by filled circles connected with a solid line and the imaginary part $(\epsilon_2)$ by empty circles connected with a dashed line. (b) Theoretical quality factors of LSPRs derived from these dielectric functions as $Q_\mathrm{LSPR}=-\epsilon_1/\epsilon_2$. (c) LSPR energy as a function of size of gallium nanoparticles, namely in-plane dipole mode in liquid \cite{10.1021/acs.jpclett.3c00094} and liquid and solid \cite{10.1021/acs.jpclett.5c02035} gallium nanoparticles on a silicon nitride membrane.}
 \label{Fig_gallium}
\end{figure*}

Bulk gallium is a metal with a melting temperature of \SI{29.7}{\celsius}. It is a non-toxic material that is acceptable to the environment \cite{10.3390/ijerph7052337, 10.1016/b978-0-444-52272-6.00474-8}. In addition, gallium has several solid-state phases, namely $\alpha$-gallium, $\beta$-gallium, $\gamma$-gallium, $\delta$-gallium, $\epsilon$-gallium, gallium-II and gallium-III, which allow for a variety of phase-changing systems \cite{10.1515/nanoph-2020-0314}. The dielectric properties of gallium are well explored by theoretical models, but there are not many measured dielectric functions. The liquid phase, $\gamma$ phase, and $\delta$ phase of gallium has a nearly Drude-like optical response from the infrared to ultraviolet spectral region, while the $\alpha$ and $\beta$ phases exhibit interband absorption in the red and green \cite{10.1002/adom.201900307, 10.1515/nanoph-2020-0314, 10.1038/s41598-017-05985-8, 10.1364/ome.9.004050}. Fortunately, this absorption is not strong enough to completely suppress plasmonic resonances \cite{10.1002/adom.201900307}. Finally, gallium offers facile and scalable preparation and good stability for nanoparticles.

Figure \ref{Fig_gallium}a shows a comparison of the experimental dielectric functions of gallium available in the literature, namely the dielectric functions of liquid gallium by Knight et al. \cite{10.1021/nn5072254} and Dogel et al. \cite{10.1103/PhysRevB.72.085403}, and solid gallium by Knight et al. \cite{10.1021/nn5072254} and McMahon et al. \cite{10.1039/c3cp43856b}. There are no significant differences in the two dielectric functions for the same phase. The major differences arise from the liquid to solid phase change. In the case of liquid gallium, the real part of the dielectric function reaches highly negative values, promising a good plasmonic activity over the entire energy range from the ultraviolet to the infrared spectral region. In the case of solid gallium, the real part of the dielectric function increases below \SI{2}{\electronvolt}, promising good plasmonic activity in the ultraviolet and visible spectral region. Figure \ref{Fig_gallium}b shows the theoretical quality factors of the LSPRs derived from these dielectric functions. In the case of liquid gallium, it increases with the energy starting around 1 at \SI{1}{\electronvolt} and reaching the value of 6 at \SI{6}{\electronvolt}. In the case of solid gallium, it increases with the energy starting around 1 at \SI{2.5}{\electronvolt} and reaching a value of 2.5 at \SI{6}{\electronvolt}. Consequently, gallium is supposed to be a suitable plasmonic platform, especially for the ultraviolet spectral region.

Gallium nanoparticles can be prepared using various bottom-up fabrication techniques such as colloidal synthesis \cite{10.1021/ja506712d, 10.1021/jacs.5c00317}, optically regulated self-assembly \cite{10.1063/1.1456260}, molecular beam epitaxy \cite{10.1063/1.2712508}, atomic beam evaporation \cite{10.1088/0957-4484/19/47/475606}, and Joule-effect thermal evaporation \cite{10.1038/s41598-020-61090-3}. Importantly, the low melting temperature of gallium allows low-temperature fabrication with low energy consumption. Despite the fact that the phase diagram of gallium nanoparticles was introduced \cite{10.1515/nanoph-2020-0314}, there are several discrepancies that need to be taken into account, as it contradicts several experimental works. First, at room temperature (\SI{25}{\celsius}) gallium nanoparticles are in the form of a supercooled liquid \cite{10.1021/acs.jpclett.3c00094, 10.1021/ja506712d, 10.1103/PhysRevLett.81.2942, 10.1021/acs.jpclett.5c02035}. Second, nanoparticles with diameters ranging from 50 to \SI{300}{\nano\meter} were found to crystallize to the $\beta$ phase of gallium with a freezing temperature around \SI{-135}{\celsius} and a metling temperature around \SI{-20}{\celsius} \cite{10.1103/PhysRevLett.81.2942, 10.1021/acs.jpclett.5c02035}. Third, nanoparticles smaller than \SI{50}{\nano\meter} were found to crystallize in the $\delta$ phase of gallium \cite{10.1021/ja506712d}. Consequently, the phase diagram for gallium nanoparticles needs to be carefully updated and reformulated with respect to the experimental results. The size of the nanoparticles needs to be reflected, as well as possible effects of the substrate and the deposition method. Finally, another challenge could be the preparation of non-spherical shapes of nanoparticles, for example, by lithography or nanomolding.

The plasmonic properties of gallium nanoparticles have been well explored, showing a wide spectral tunability from the ultraviolet to the near-infrared spectral region at room temperature \cite{10.1021/nn5072254, 10.1021/acs.jpclett.3c00094, 10.1002/smll.201902920}. Furthermore, the impact of the liquid to solid phase transition on the plasmonic properties of gallium nanoparticles has been explored, while differences in the LSPR energies between liquid gallium and $\beta$-gallium nanoparticles are minor \cite{10.1021/acs.jpclett.5c02035}. As a result, the energy shift related to the temperature-induced phase change is relatively small to be utilized, for example, as a temperature sensor. The performance of gallium nanoparticles is, in the case of temperature-dependent experiments, unaffected by the liquid to solid phase change of gallium, which is valuable for cryogenic temperature-suppressed non-radiative recombination in surface-enhanced Raman spectroscopy \cite{10.1364/OL.447995, 10.1364/OE.495426}. In addition, plasmonic properties of Ga-\ce{Ga2O3} core-shell structures \cite{10.1088/1361-6528/aa8505, 10.1088/1361-6528/aacb91} and gallium-indium \cite{10.1038/s41598-019-41789-8} and gallium-silver \cite{10.1063/1.4906950} alloys have been reported.

Figure \ref{Fig_gallium}c shows the dipole LSPR energy as a function of the size of gallium nanoparticles available in the literature, namely the in-plane dipole mode in liquid \cite{10.1021/acs.jpclett.3c00094} and liquid and solid \cite{10.1021/acs.jpclett.5c02035} lens-shaped gallium nanoparticles on a silicon nitride membrane. The results show a wide tunability of gallium nanostructures from the ultraviolet region for the size of structures below \SI{50}{\nano\meter} to the visible spectral region for the size of structures up to \SI{200}{\nano\meter}. In addition, the liquid gallium nanostructures above \SI{200}{\nano\meter} are intended to represent the plasmonic platform for the near-infrared spectral region.

There are numerous applications for such nanoparticles, including DNA biosensing platforms \cite{10.1039/C6NR00926C}, luminescence enhancement of \ce{MoS2} monolayers \cite{10.1039/C8NA00094H}, and surface-enhanced Raman spectroscopy applications \cite{10.1021/ja903321z, 10.1021/nl401145j, 10.1007/s10854-021-05566-6, 10.1021/acs.jpcc.1c10426}. Further research in gallium-based plasmonics could include applications that utilize the room- to cryo-temperature stability of LSPR in gallium nanoparticles, Ga-\ce{GaN} core-shell structures, or other alloys such as gallium-aluminium.

\section{Magnesium}

\begin{figure*}[ht]
 \centering
 \includegraphics[width=16cm]{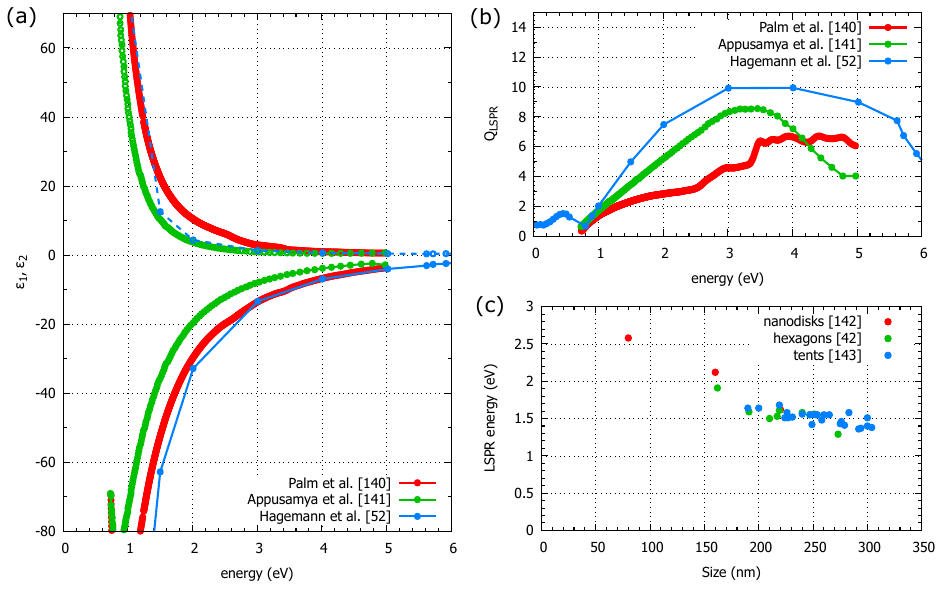}
 \caption{Magnesium plasmonics: (a) Experimental dielectric functions of magnesium by Palm et al. \cite{10.1021/acsphotonics.8b01243}, Appusamya et al. \cite{10.1016/j.mseb.2013.11.009}, and Hagemann et al. \cite{10.1364/JOSA.65.000742}. The real part $(\epsilon_1)$ is plotted by filled circles connected with a solid line and the imaginary part $(\epsilon_2)$ by empty circles connected with a dashed line. (b) Theoretical quality factors of LSPRs derived from these dielectric functions as $Q_\mathrm{LSPR}=-\epsilon_1/\epsilon_2$. (c) LSPR energy as a function of size of magnesium nanostructures, namely magnesium nanodisks \cite{10.1021/acs.nanolett.5b03029}, hexagons \cite{10.1021/acs.nanolett.8b00955}, and tent-shaped structures \cite{10.1021/acsnano.0c01427}.}
 \label{Fig_magnesium}
\end{figure*}

Magnesium is the sixth most abundant element on Earth. Bulk magnesium is a metal with a melting temperature of \SI{650}{\celsius}. Magnesium is also a very important biogenic element. It is found in all green plants, where it is a component of chlorophyll. It is also one of the important biogenic elements in animal organisms. It is highly reactive and elemental magnesium is a strong reducing agent. It has a higher plasmonic quality factor than aluminium across the visible spectral region, making it an attractive framework for plasmonics. The hexagonal, folded, and kite-shaped shapes were theoretically expected from a modified Wulff construction for single crystal and twinned magnesium nanostructures, and their plasmonic properties have been introduced, highlighting the ability of magnesium to sustain LSPRs in the ultraviolet, visible, and near-infrared spectral region \cite{10.1021/acs.jpcc.0c03871}.

Figure \ref{Fig_magnesium}a shows a comparison of three experimental dielectric functions of magnesium available in the literature, namely by Palm et al. \cite{10.1021/acsphotonics.8b01243}, Appusamy et al. \cite{10.1016/j.mseb.2013.11.009}, and Hagemann et al. \cite{10.1364/JOSA.65.000742}. There seem to be no significant differences. However, they may still influence the simulation results, so the dielectric function should be chosen carefully. The real part of the dielectric function reaches highly negative values in the infrared and visible spectral range, promising good plasmonic activity. Figure \ref{Fig_magnesium}b shows the theoretical quality factors of the LSPRs derived from these dielectric functions. They reach values between 3 and 10 in the visible and ultraviolet spectral regions. Consequently, magnesium is a suitable plasmonic platform in the ultraviolet and visible spectral region. 

Magnesium nanostructures can be chemically synthesized \cite{10.1021/acs.nanolett.8b00955,10.1021/acsnano.0c01427} as well as manufactured by colloidal hole-mask lithography \cite{10.1021/acs.nanolett.5b03029} or nanosphere lithography \cite{10.1063/5.0210650}. The stability of plasmonic magnesium nanoparticles can be improved by encapsulation in a polydopamine shell \cite{10.1039/d1nr06139a}. Full modal analysis of LSPR magnesium hexagons \cite{10.1021/acs.nanolett.8b00955} and tent-shaped structures \cite{10.1021/acsnano.0c01427} was performed.

Figure \ref{Fig_magnesium}c shows the dipole LSPR energy as a function of the size of magnesium nanoparticles available in the literature, namely magnesium nanodisks \cite{10.1021/acs.nanolett.5b03029}, hexagons \cite{10.1021/acs.nanolett.8b00955}, and tent-shaped structures \cite{10.1021/acsnano.0c01427}. The results show a relatively narrow tunability of magnesium nanostructures from the visible region for the size of structures between 50 and \SI{200}{\nano\meter} to the near-infrared spectral region for the larger structures. In addition, smaller magnesium nanostructures are intended to represent the plasmonic platform for the ultraviolet spectral region.

The potential of magnesium based nanostructures for surface-enhanced Raman spectroscopy is demonstrated with enhancement factors of around 100 for pure magnesium and around 1000 for magnesium with palladium \cite{10.1021/acsnano.4c06858}. Magnesium is more efficient at converting light into heat than gold at near-infrared wavelengths, so magnesium nanoparticles can be used as inexpensive and biodegradable photothermal platforms \cite{10.1021/acs.nanolett.3c03219}. In addition, magnesium nanoparticles have been introduced in biomedical applications as a platform for hydrogen cancer therapy \cite{10.1016/j.chempr.2022.07.001}. Magnesium nanohelices were studied as chiral nanoparticles having a remarkable chiroptical effect in the ultraviolet region together with an enhanced LSPR sensitivity \cite{10.1039/c6cc06800f}. A magnesium plasmonic device whose transmittance can be tuned by water can serve as an optical sensor to measure ambient humidity and monitor changes in humidity over time \cite{10.1007/s12274-018-2028-6}. Finally, palladium-magnesium bimetallic nanocomposites were shown to be promising highly selective hydrogenation catalysts under conventional thermally-driven conditions \cite{10.1039/d3nr00745f}.

Moreover, magnesium nanostructures can be used as a platform for dynamic plasmonics. It is based on a reversible phase transition between the metallic state of pure magnesium (Mg) and the dielectric state of magnesium hydride (\ce{MgH2}) through hydrogenation and dehydrogenation under hydrogen (\ce{H2}) and oxygen (\ce{02}) atmosphere, respectively \cite{10.1021/acs.nanolett.5b03029, 10.1021/acs.accounts.9b00157}. Alternatively, this phase transition may be driven by hydrogenation in the hydrogen atmosphere and dehydrogenation by thermal desorption of hydrogen in the argon atmosphere at \SI{95}{\celsius} \cite{10.1021/acs.nanolett.8b01277}.

\section{Tin}

\begin{figure*}[ht]
 \centering
 \includegraphics[width=16cm]{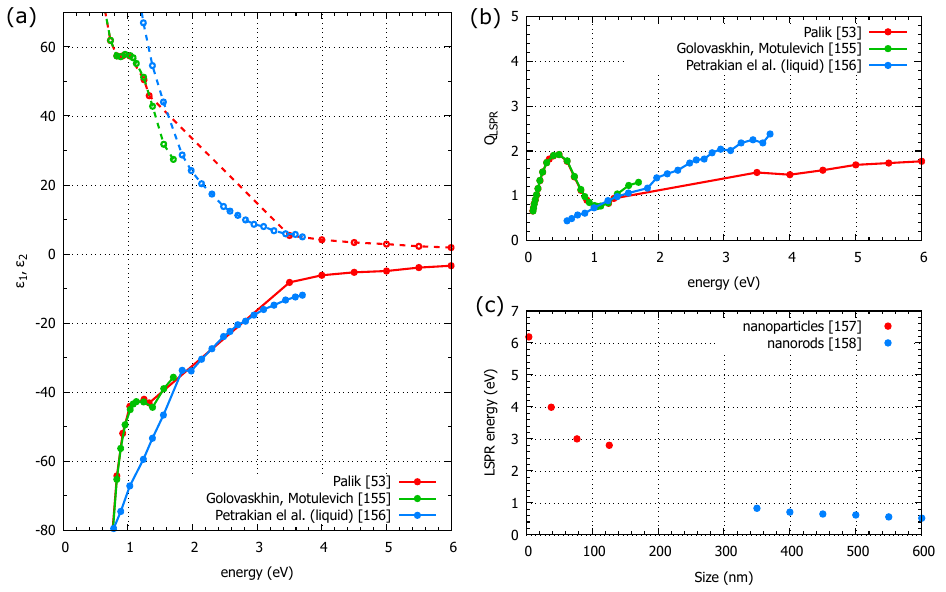}
 \caption{Tin plasmonics: (a) Experimental dielectric functions of tin by Palik \cite{10.1016/c2009-0-20920-2}, Golovashkin and Motulevich \cite{e_019_02_0310}, and for the liquid tin by Petrakian et al. \cite{10.1103/PhysRevB.21.3043}. The real part $(\epsilon_1)$ is plotted by filled circles connected with a solid line and the imaginary part $(\epsilon_2)$ by empty circles connected with a dashed line. (b) Theoretical quality factors of LSPRs derived from these dielectric functions as $Q_\mathrm{LSPR}=-\epsilon_1/\epsilon_2$. (c) LSPR energy as a function of size of tin nanostructures, namely tin sphecical nanoparticles \cite{10.1021/acs.jpcc.8b10851} and tin nanorods on glass \cite{10.1007/s11468-024-02505-z}.}
 \label{Fig_tin}
\end{figure*}

Tin is a soft, malleable, and ductile metal. The melting temperature of the bulk tin is \SI{232}{\celsius}. However, the melting point is reduced to \SI{150}{\celsius} for nanoparticles below \SI{10}{\nano\meter} \cite{10.1103/PhysRevLett.77.99, 10.1088/0957-4484/22/22/225701}. The surface melting of a single tin nanoparticle with a diameter of \SI{47}{\nano\meter} was observed at 200--\SI{225}{\celsius} by in situ transmission electron microscopy showing a wide hysteresis, as the crystallization of the liquid tin nanoparticle occurred with a large overcooling at \SI{100}{\celsius} \cite{10.1021/acs.nanolett.3c00943}. The wide hysteresis (around \SI{130}{\celsius}) of the solid-liquid phase change of the tin nanoparticles was determined by optical excitation measurements of the localized surface plasmon resonance in the tin nanoparticles \cite{10.1021/nl100044k}. Moreover, the local thermal expansion coefficient of single liquid tin nanoparticles was evaluated \cite{10.1038/s41598-025-88496-1}. In addition, tin has several solid-state phases. $\beta$-tin (also called white tin) is the allotrope of tin that is stable at and above room temperature. It is metallic with a body-centered tetragonal crystal structure. In cold $\beta$-tin tends to spontaneously transform into $\alpha$-tin, which is a phenomenon known as the tin pest. $\alpha$-tin (or gray tin) is a nonmetallic form with a diamond cubic crystal structure that is stable below \SI{13}{\celsius}. At temperatures above \SI{160}{\celsius} and pressures above several gigapascals $\gamma$-tin and $\sigma$-tin are reported \cite{10.1007/bf02755923}.

Figure \ref{Fig_tin}a shows a comparison of three experimental dielectric functions of tin available in the literature, namely by Palik \cite{10.1016/c2009-0-20920-2}, Golovashkin and Motulevich \cite{e_019_02_0310}, and for the liquid tin by Petrakian et al. \cite{10.1103/PhysRevB.21.3043}. There seem to be no significant differences. However, all of these dielectric functions suffer from their narrow spectral range and low number of data points. The real part of the dielectric function reaches highly negative values, promising plasmonic activity over the entire energy range from the ultraviolet to the infrared spectral region. Figure \ref{Fig_tin}b shows the theoretical quality factors of the LSPRs derived from these dielectric functions. They reach values between 1 and 2, predicting that tin is not a perfect plasmonic material. However, further experiments are required to obtain a more accurate dielectric function of tin.

Tin nanostructures can be synthesized chemically \cite{10.1021/acs.jpcc.8b10851}, as well as manufactured by electron beam lithography \cite{10.1007/s11468-012-9445-2} or sequential self-assembly (chemical dealloying) \cite{10.1002/adom.202300568, 10.1002/adom.202403432}. The plasmonic properties of tin nanoparticles have been thoroughly explored, including a size-dependent evolution of plasmonic modes and surface-enhanced Raman spectroscopy performance of $\beta$-tin nanoparticles \cite{10.1021/acs.jpcc.8b10851} and tuning the plasmon resonance of metallic tin nanocrystals dispersed in silicon dioxide and amorphous silicon matrix \cite{10.1007/s00339-010-5805-y}. Moreover, the optical transmission of hexagonal oriented tin nanobars was discussed to create a plasmonic meatsurface for the near-infrared spectral region \cite{10.1007/s11468-024-02505-z}. Tin disks with a diameter of \SI{155}{\nano\meter} and height of \SI{50}{\nano\meter} reveal a transmission peak around \SI{2.3}{\electronvolt} and their sensing possibilities have been discussed \cite{10.1007/s11468-012-9445-2}.

Figure \ref{Fig_tin}c shows the dipole LSPR energy as a function of the size of tin nanoparticles available in the literature, namely tin spherical nanoparticles \cite{10.1021/acs.jpcc.8b10851} and tin nanorods on glass \cite{10.1007/s11468-024-02505-z}. The results show a wide tunability of the tin nanostructures from the ultraviolet region for the size of structures below \SI{100}{\nano\meter} to the near-infrared spectral region for structures of length between 300 and \SI{600}{\nano\meter}. In addition, tin nanostructures with a size between 100 and \SI{250}{\nano\meter} are intended to represent the plasmonic platform for the visible spectral region.

The coupling of localized surface plasmons was explored in tin nanoparticles dispersed in a copper matrix \cite{10.1002/adom.202300568}. The Janus nanoparticle was effectively formed by introducing a silver nanoantenna active in the visible light spectral region on one side of a tin nanoparticle with localized surface plasmons located in the ultraviolet range \cite{10.1002/adom.202403432}. In addition, the Begrenzung effect that leads to the suppression of surface plasmons was discussed for tin nanoparticles encapsulated in silicon nitride \cite{10.1103/PhysRevB.110.205411}. Tin nanoparticles have been proposed as a relevant additive to the photoactive silicon layer in plasmon assisted silicon solar cells, as tin is in the same group IV as silicon and therefore does not introduce any additional energy level into the band gap nor create recombination centers \cite{10.1002/pssr.201004190}.

\section{Other non-noble metals}

The following non-noble metals, antimony (Sb), chromium (Cr), indium (In), lead (Pb), molybdenum (Mo), nickel (Ni), potassium (K), selenium (Se), sodium (Na), tellurium (Te), titanium (Ti), tungsten (W), and zinc (Zn), are less frequently used in plasmonic applications. Therefore, we will discuss them briefly below. The rationale underlying this concise synopsis is chiefly the circumscribed extent of general understanding. Finally, we note that the list does not contain all non-noble metals, so there is still a possibility to pioneer the plasmonic properties of some of the unexplored non-noble metals.

\subsection{Antimony}
The first non-noble metal in this section is antimony. Tunable localized surface plasmon resonances have been reported in biocompatible antimony nanopolyhedrons with high photothermal conversion efficiency and good photothermal stability \cite{10.1002/adma.202100039}.

\subsection{Chromium}
The weak plasmonic activity of chromium was theoretically predicted \cite{10.1021/jp405773p}, but there are only a few applications in the literature that include chromium microrods \cite{10.1088/1361-6528/aca339} and chromium nitride (Cr$_2$N) nanoparticles \cite{10.1063/5.0109806}.

\subsection{Indium}
In the case of indium, the field of applications is slightly larger. Indium nanoparticles were found to be a promising candidate for deep-ultraviolet plasmonics \cite{10.1063/5.0271244, 10.1039/d4cp00095a, 10.1039/c4cp05743k}. Indium nanocrystals were also used for the selective enhancement of blue upconversion luminescence \cite{10.1002/adfm.201901242}, to improve the efficiency of silicon solar cells \cite{10.1016/j.apsusc.2020.145275}, and for surface-enhanced Raman scattering \cite{10.1039/c7ra03317f}.

\subsection{Lead}
The plasmonic properties of lead nanoparticles have not been explored in detail. However, the plasmonic response of lead films deposited on nanostructured substrates was investigated, showing that lead can be used as an alternative plasmonic material within the visible to near-infrared spectral range \cite{10.1063/5.0016131}.

\subsection{Molybdenum}
Localized surface plasmon resonances were introduced for molybdenum microstructures \cite{10.3390/photonics11100950}. However, with respect to their size and spatial distribution, it is disputable whether the observed effect can be assigned to the localized surface plasmon resonance or not. As a result, there has been no systematic study of plasmon resonances in individual molybdenum nanostructures. In addition, tunable plasmonic resonance was observed in molybdenum oxides (MoO$_3$ and MoO$_{3-x}$) \cite{10.1002/adfm.201806699, 10.1002/adom.202301821} and a strong surface plasmon resonance in molybdenum nitride (MoN) nanosheets has been introduced as a promising surface enhanced Raman scattering platform \cite{10.1038/s41467-020-17628-0}

\subsection{Nickel}
In the case of nickel, its plasmonic properties, which have been thoroughly explored \cite{10.1039/c7cp01769c}, can be combined with its magnetic properties. In this sense, plasmon-induced demagnetization and magnetic switching in nickel nanoparticle arrays was introduced \cite{10.1063/1.5012857}.

\subsection{Potassium}
Potassium plasmonic nanostructures were fabricated using a thermo-assisted nanoscale embossing technique to create nanostructures with varying periodicities that support high-quality surface plasmon modes \cite{10.1021/acs.nanolett.3c02054}. In addition, sodium-potassium liquid alloys have been discussed \cite{10.1021/acs.nanolett.3c02054}. However, there is no systematic study of localized surface plasmons in individual potassium nanostructures.

\subsection{Selenium}
The plasmonic activity has been reported for selenium nanoparticles that revealed a significant surface plasmon resonance peak at \SI{4.7}{\electronvolt}, and their protective effect against chromium tubular necrosis has been discussed \cite{10.1007/s11468-024-02539-3}.

\subsection{Sodium}
Sodium is predicted to be an ideal plasmonic material with very low optical loss across visible to near-infrared. A scalable fabrication method for sodium nanostructures was developed by combining phase-shift photolithography and a thermo-assisted spin-coating process. Using this method, sodium nanopit arrays were manufactured with varying periodicity that revealed tunable surface plasmon polariton modes ranging from visible to near-infrared \cite{10.1021/acs.nanolett.2c03643}. In addition, hot electron dynamics has been investigated in nanostructured sodium thin films on polyurethane supports, showing a unique early time response that provides key information on sodium-based plasmonics \cite{10.1021/acsnano.5c04946}. However, there is no systematic study of localized surface plasmons in individual sodium nanostructures.

\subsection{Tellurium}
Tellurium is the first reported material that simultaneously has plasmonic-like and all-dielectric properties in the solar radiation region. Tellurium nanoparticles with a wide size distribution were demonstrated to absorb more than 85\% solar radiation throughout the spectrum, so it can be expected to be an advanced photothermal conversion material for solar-enabled water evaporation \cite{10.1126/sciadv.aas9894}.

\subsection{Titanium}
The weak plasmonic activity of titanium was theoretically predicted \cite{10.1021/jp405773p}, but in the literature no applications have been reported for pure titanium nanostructures. However, titanium nitride (TiN) is considered an alternative ceramic platform for plasmonic applications \cite{10.1364/ome.2.000478, 10.1039/d3ma00965c}.

\subsection{Tungsten}
Characteristics of localized surface plasmons of tungsten nanowires with different diameters and lengths have been explored \cite{10.1088/2053-1591/aca5ee}, but no direct application has been reported. However, tungsten oxides, namely WO$_{3-x}$ with abundant oxygen vacancies \cite{10.1039/c9nr03741a, 10.1021/acs.chemmater.4c02233} and W$_{18}$O$_{49}$ modified with HCl \cite{10.1002/adma.202404738}, have been used as nonmetallic plasmonic catalysts.

\subsection{Zinc}
It has been experimentally and numerically confirmed that silica-dispersed zinc nanoparticles exhibit two optical extinction peaks that were ascribed to surface plasmon resonances in the broad sense \cite{10.1088/0957-4484/18/39/395707, 10.1088/1361-6528/aa950d, 10.1134/S1087659619030155}. However, a comprehensive study of localized surface plasmon resonances is lacking. Furthermore, we note that zinc oxide (ZnO) nanoparticles represent a commonly used catalytic system \cite{10.1021/acscatal.8b04105, 10.1021/acscatal.8b04873} that can be modified with noble metals to further enhance its properties \cite{10.1039/c4ra03158j, 10.1038/s41929-021-00708-9}. 

\section*{Conclusions}

In summary, this review of the literature dealing with the investigation of localized surface plasmons in non-noble metal nanoparticles provides an overview of the plasmonic properties of non-noble metals. We focused mostly on widely used non-noble plasmonic metals such as aluminium (Al), bismuth (Bi), copper (Cu), gallium (Ga), magnesium (Mg), and tin (Sn). In addition, the following less frequently used plasmonic non-noble metals are discussed, too: antimony (Sb), chromium (Cr), indium (In), lead (Pb),  molybdenum (Mo), nickel (Ni), potassium (K), selenium (Se), sodium (Na), tellurium (Te), titanium (Ti), tungsten (W), and zinc (Zn). We compared the experimental dielectric functions and the theoretical quality factors of localized surface plasmon resonances calculated using them. In addition, we reviewed the spectral tunability of real nanostructures by plotting the dipole mode energy as a function of their size. Finally, examples of applications were briefly summarized for every metal.  

Aluminium, bismuth, gallium, magnesium, and tin nanostructures offer wide tunability of the dipole mode with increasing their size from the ultraviolet region through the visible spectral region to the near-infrared spectral region. In addition, magnesium stands out as a chemically switchable plasmonic platform based on hydrogenation and dehydrogenation. Copper is supposed to be an excellent plasmonic platform for the near-infrared and red part of visible spectral region with an excellent theoretical quality factor of localized surface plasmon resonances reaching values of 20 to 45.

Our summary therefore compared the plasmonic properties of non-noble metals and briefly introduced their potential to the readers. Nevertheless, the plasmonics of non-noble metals is still an open field with a huge potential for new discoveries. These may include a systematic study of less explored non-noble metals, introduction of any other non-noble metal to the plasmonic family, or development of new applications and fabrication protocols for large-scale devices.

\begin{acknowledgement}

This work is supported by the OP JAK Excellent Research program (project QM4ST, No. CZ.02.01.01/00/22\_008/0004572), and Brno University of Technology (project No. FSI-S-23-8336). M.F. acknowledges the support of the Brno Ph.D. talent scholarship.

\end{acknowledgement}

\bibliography{literatura}

\end{document}